\documentclass[11pt]{article}
\usepackage{hyperref}
\pdfoutput=1

\begin{document}
\title{Flame Reconstruction Using Synthetic Aperture Imaging}
\author{Preston Murray, Jonathon Pendlebury, Dale Tree and Tadd Truscott \\
\\\vspace{6pt} Department of Mechanical Engineering
\\ Brigham Young University, Provo, UT 84602, USA}
\maketitle

\begin{abstract}

Flames can be formed by burning methane (CH$_4$). When oxygen is scarce, carbon particles nucleate into solid particles called soot. These particles emit photons, making the flame yellow. Later, methane is pre-mixed with air forming a blue flame; burning more efficiently, providing less soot and light. 

Imaging flames and knowing their temperature are vital to maximizing efficiency and validating numerical models.  Most temperature probes disrupt the flame and create differences leading to an inaccurate measurement of the flame temperature.  We seek to image the flame in three dimensions using synthetic aperture imaging. This technique has already  successfully   measured velocity fields of a vortex ring \cite{Belden2010}. 

Synthetic aperture imaging is a technique that views one scene from multiple cameras set at different angles, allowing some cameras to view objects that are obscured by others.  As the resulting images are overlapped different depths of the scene come into and out of focus, known as focal planes, similar to tomography. These focal planes can be used to extract three-dimensional information about the scene. This procedure was used to extract the outer edge of a oxygen-starved methane flame (yellow) from which the three-dimensional flame was reconstructed. When the reconstructed image was compared to the raw image from the central camera the two shapes corresponded well. This experiment in the fluid dynamics  video (entry \#: V041) demonstrated that a three-dimensional flame can be reconstructed by combining images from multiple cameras using synthetic aperture imaging. 
\end{abstract}

\bibliographystyle{abbrv}
\bibliography{bib}

\begin{thebibliography}{1}

\bibitem{Belden2010}
J.~Belden, T.~Truscott, M.~Axiak, and A.~Techet.
\newblock Three-dimensional synthetic aperture particle image velocimetry.
\newblock {\em Measurement Science and Technology}, 21, September 2010.

\end{thebibliography}

\end{document}